\documentclass[10pt, a4paper]{article}
\usepackage[margin=1in]{geometry}
\usepackage{amsmath,amssymb,amsthm}
\usepackage{array}
\usepackage[shortlabels]{enumitem}
\usepackage{graphicx}

\usepackage[round]{natbib}
\usepackage{authblk}

\newtheorem{thm}{Theorem}
\newtheorem{lem}{Lemma}

\newtheorem{obs}{Observation}

\newtheorem{claim}{Claim}

\newcommand{\ceil}[1]{\left\lceil #1 \right\rceil}

\newcommand{\istar}{i^\star}

\newcommand{\E}{\mathbb{E}}

\newcommand{\Tpred}{\widehat{T}}
\newcommand{\Jpred}{\widehat{J}}

\newcommand{\A}{\mathcal{A}}

\newcommand{\con}{\chi}

\DeclareMathOperator{\sol}{SOL}
\DeclareMathOperator{\opt}{OPT}

\newcommand{\yrcite}[1]{(\citeyear{#1})}
\renewcommand{\cite}[1]{\citep{#1}}

\title{Improved Learning-Augmented Algorithms for the Multi-Option Ski Rental Problem via Best-Possible Competitive Analysis}
\author[]{Yongho Shin}
\author[]{Changyeol Lee}
\author[]{Gukryeol Lee}
\author[]{Hyung-Chan An\thanks{Corresponding author: \texttt{hyung-chan.an@yonsei.ac.kr}}}
\affil[]{Department of Computer Science, Yonsei University, Seoul, South Korea}
\date{}

\begin{document}

\maketitle
\begin{abstract}
In this paper, we present improved learning-augmented algorithms for the multi-option ski rental problem. Learning-augmented algorithms take ML predictions as an added part of the input and incorporates these predictions in solving the given problem. Due to their unique strength that combines the power of ML predictions with rigorous performance guarantees, they have been extensively studied in the context of online optimization problems. Even though ski rental problems are one of the canonical problems in the field of online optimization, only deterministic algorithms were previously known for multi-option ski rental, with or without learning augmentation. We present the first randomized learning-augmented algorithm for this problem, surpassing previous performance guarantees given by deterministic algorithms. Our learning-augmented algorithm is based on a new, provably best-possible randomized competitive algorithm for the problem. Our results are further complemented by lower bounds for deterministic and randomized algorithms, and computational experiments evaluating our algorithms' performance improvements.
\end{abstract}

\section{Introduction}
A \emph{learning-augmented algorithm} takes an ML prediction as an added part of the input and incorporates this prediction in solving the given problem.
One major advantage of these algorithms is that they can benefit from the powerful ML predictions while  yielding provable performance guarantees at the same time. These guarantees often surpass those given by classical algorithms without predictions, and they are obtained with no assumptions on how the ML predictions are generated, keeping their black-box natures.
The recent success of learning-augmented algorithms is particularly remarkable in the field of \emph{online optimization}, where we are given the input in an online manner over multiple timesteps and forced to make irrevocable decisions at each timestep.
It is not at all surprising that learning-augmented algorithms are useful in this setting, because the challenge in designing a classical competitive algorithm is usually in avoiding the worst-case without any knowledge on the future input, where ML predictions can serve as a substitute for this knowledge.
The success of learning-augmented algorithms in online optimization is evidenced by the seminal work of Lykouris and Vassilvitskii~\yrcite{lykouris2021competitive} and subsequent studies including~\cite{kumar2018improving, bamas2020primal, lattanzi2020online} for example.

\emph{Ski rental problems} are one of the canonical problems in online optimization that have been extensively studied, with or without learning augmentation.
They succinctly capture the key nature of online optimization, and many algorithmic techniques in online optimization have been devised from their studies.
For learning-augmented algorithms as well, ski rental problems naturally have been serving as an important testbed~\cite{kumar2018improving, gollapudi2019online, bamas2020primal, wei2020optimal}.

Whilst the ski rental problem emerges in a variety of applications~\cite{fleischer2001bahncard, karlin2001dynamic, meyerson2005parking},
it is perhaps easiest to state it as a simple problem of renting skis (hence the name) as follows.
In the \emph{multi-option ski rental problem}, we are given a set of $n$ renting options: for $i = 1, \cdots, n$, we have the option of renting skis for $d_i \in \mathbb{Z}_+ \cup \{\infty\}$ days at cost $c_i \in \mathbb{Q}_+$.
Let $T$ be the number of days we will go skiing, but it depends on, say, the weather, so we do not know $T$ in advance; we will learn that a specific day was indeed the last day of skiing only at the end of that day. The objective of this problem is to ensure that we rent skis for the whole $T$ days, while minimizing the total renting cost.
Traditionally, the \emph{rent-or-buy} case, where we only have  two options $(d_1,c_1)=(1,1)$ and $(d_2,c_2)=(\infty,B)$,
was extensively studied, both under the classical competitive analysis setting without learning augmentation~\cite{karlin1988competitive, karlin1994competitive, buchbinder2007online} and the learning-augmented settings~\cite{kumar2018improving, gollapudi2019online, angelopoulos2020online, bamas2020primal, banerjee2020improving, wei2020optimal}.
The general multi-option problem has been also studied in both settings~\cite{zhang2011ski, ai2014multi, wang2020online, anand2021regression}.

In this paper, we present the first \emph{randomized} learning-augmented algorithm for the multi-option ski rental problem.
Previously, Anand et al.~\yrcite{anand2021regression} gave a  deterministic learning-augmented algorithm for the same problem. The performance guarantee of their algorithm is stated under the standard consistency-robustness scheme: their algorithm is $(1+\lambda)$-consistent and $(5+5/\lambda)$-robust, where $\lambda$ is the trade-off parameter (or the ``level of confidence'') that dictates how much the algorithm would ``trust'' or ``ignore'' the prediction and determines the performance guarantees.\footnote{Intuitively speaking, consistency is a performance guarantee that is valid only when the ML prediction is accurate, whereas robustness is always valid. Both are measured as a worst-case ratio of the algorithm's output to the true optimum with hindsight; see Section~\ref{sec:prelim} for the formal definitions.}
Their learning-augmented algorithm was obtained by modifying their classical competitive algorithm to use the given ML prediction~\cite{anand2021regression}; hence, in order to attain an improved learning-augmented algorithm, it is natural to delve in devising a more competitive algorithm for the classical problem.

Section~\ref{sec:det} is a warm-up where we first consider deterministic algorithms. We present a competitive algorithm without learning augmentation in Section~\ref{sec:det-comp}, which is very similar to Anand et al.'s algorithm and even has the same competitive ratio. We nonetheless present this algorithm as the subtle difference turns out to be useful in obtaining the improved learning-augmented algorithm presented in Section~\ref{sec:det-pred}. To obtain this improvement, we allow our learning-augmented algorithm to ```ignore'' the prediction. The previous algorithm~\cite{anand2021regression} internally computes the optimal solution assuming the (scaled) prediction is accurate, and insists on including this in the algorithm's output as a means of guaranteeing the consistency. On the other hand, our algorithm  is capable of choosing not to do so when the trade-off parameter says the ML prediction is not too reliable. While this may sound natural, it is a new characteristic of our algorithm that leads to the improvement in the performance trade-off.

In Section~\ref{sec:rand}, we show how randomization can improve our algorithms. Section~\ref{sec:rand-comp} presents an $e$-competitive algorithm for the multi-option ski rental problem. In the doubling scheme employed by Anand et al.~\yrcite{anand2021regression}, one can adversarially construct a malicious instance by calculating the doubling budgets the algorithm will use.
We can prevent the adversary from this exploitation by randomizing the budgets; our algorithm reveals that only a small amount of randomness suffices to obtain an $e$-competitive algorithm, which is provably the best possible (cf. Section~\ref{sec:lbound-rand}).
In Section~\ref{sec:rand-pred}, we base on this $e$-competitive algorithm to improve the performance trade-off of our learning-augmented algorithm. Our learning-augmented algorithm is $\chi(\lambda)$-consistent and $(e^\lambda/\lambda)$-robust, where
\[
\chi(\lambda) := \begin{cases}
1 + \lambda, & \text{if } \lambda < \frac{1}{e}, \\
(e + 1) \lambda - \ln \lambda - 1, & \text{if } \lambda \geq \frac{1}{e}
,\end{cases}
\]improving over the previous algorithms.

In Section~\ref{sec:lbound}, we present lower bounds.
We first propose an auxiliary problem called the \emph{button problem} that is more amenable to lower-bound arguments, which we then reduce to the multi-option ski rental problem.
In Section~\ref{sec:lbound-rand}, we show that our algorithm in Section~\ref{sec:rand-comp} is the best possible: i.e., for all constant $\rho<e$, there does not exist a randomized $\rho$-competitive algorithm for the multi-option ski rental problem. To prove this, we carefully formulate a linear program (LP) that bounds the competitive ratio of any randomized algorithms for a given instance. We then obtain the desired lower bound by analytically constructing feasible solutions to the dual of this LP.
Section~\ref{sec:lbound-det} then begins with showing a lower bound of $4$ on the competitive ratio of deterministic algorithms. Although Zhang et al.~\yrcite{zhang2011ski} already gives the same lower bound of $4$, we still present this proof to extend it into a lower bound for deterministic learning-augmented algorithms in Section~\ref{sec:lbound-det}. 

Lastly, in Section~\ref{sec:exp}, we experimentally evaluate the performance of our learning-augmented algorithms. We conduct computational experiments to measure the performance of our algorithms under a similar setting to~\cite{kumar2018improving}, to demonstrate the performance improvements our algorithms bring.

\paragraph{Related Work}
Recently, learning-augmented algorithms have been  studied for a broad range of traditional optimization problems. Many studies introduce ML predictions to online optimization problems in particular, including caching~\cite{rohatgi2020near, lykouris2021competitive, im2022parsimonious}, matching~\cite{antoniadis2020secretary, lavastida2021learnable}, and graph/metric problems~\cite{antoniadis2020online, azar2022online, shaofeng2022tenth} for example. We refer interested readers to the survey of Mitzenmacher and Vassilvitskii~\yrcite{mitzenmacher2022algorithms} for a more thorough review.

Being a traditional online optimization problem itself, the ski rental problem is no exception and is widely studied. Karlin et al.~\yrcite{karlin1994competitive} presented a randomized $(e/(e-1))$-competitive algorithm for the rent-or-buy problem, which is the best possible. Zhang et al.~\yrcite{zhang2011ski} presented a deterministic $4$-competitive algorithm for the multi-option ski rental problem under the decreasing marginal cost assumption. The problem has also been studied in a variety of settings, including the multi-shop ski rental problem~\cite{ai2014multi}, the Banchard problem~\cite{fleischer2001bahncard}, the dynamic TCP acknowledgement problem~\cite{karlin2001dynamic}, and the parking permit problem~\cite{meyerson2005parking} for example.

\section{Preliminaries} \label{sec:prelim}
In the \emph{multi-option ski rental problem without learning augmentation}, we are given a set of $n$ renting options: each option~$i$ covers $d_i \in \mathbb{Z}_+ \cup \{\infty\}$ days at cost $c_i \in \mathbb{Q}_+$. A solution to this problem is a sequence of options. In this problem, at the beginning of each skiing day, if the day is not yet covered by our solution, we must choose an option and add it to the ``current'' output solution. Let $T$ be the number of skiing days, which is revealed only at the end of day $T$. The goal of the problem is to cover these $T$ days by a solution of minimum cost. We say an algorithm is \emph{$\gamma$-competitive} if the expected cost of the algorithm's output is no greater than $\gamma$ times the minimum cost.

In the \emph{learning-augmented multi-option ski rental problem}, we are additionally given a prediction $\Tpred$ on the number of skiing days $T$. The performance of a learning-augmented algorithm is measured by the standard consistency-robustness analysis. That is, we say that an algorithm is \emph{$\chi$-consistent} if the algorithm satisfies $\E[\sol] \leq \chi \cdot \opt(\Tpred)$ when the prediction is accurate (i.e., $\Tpred=T$), and the algorithm is \emph{$\rho$-robust} if $\E[\sol] \leq \rho \cdot \opt(T)$ for all $T$ regardless how accurate the prediction $\Tpred$ is.

Given two solutions $\sol_1$ and $\sol_2$, we say that we \emph{append} $\sol_2$ to $\sol_1$ when we add the options in $\sol_2$ to the end of $\sol_1$. For example, if $\sol_1$ is to choose option 1 and $\sol_2$ is to choose option 2, option 3, and option 3, adding $\sol_2$ to $\sol_1$ yields a solution that chooses option 1, option 2, option 3, and option 3. 

For each $t \in \mathbb{Z}_+$, let $\opt(t)$ be a minimum-cost solution that covers (at least) $t$ days. Ties are broken arbitrarily. We may slightly abuse the notation and write $\opt(t)$ to denote its cost rather than the solution itself when clear from the context. Without loss of generality, let us assume that $\opt(1) \geq 1$; otherwise, we may divide the cost of every option by $\opt(1)$. In the rest of this paper, algorithms would append an optimal solution $\opt(t)$ for some $t$ to the ``current'' solution. Note that this $\opt(t)$ can be computed for any $t$ without knowing the true number of skiing days $T$: a standard dynamic programming technique can be used to obtain $\opt(t)$.

\section{Deterministic Algorithms} \label{sec:det}
In this section, we present our deterministic algorithms for the multi-option ski rental problem. We begin with a 4-competitive algorithm in Section~\ref{sec:det-comp}. Although this algorithm is very similar to Anand et al.'s algorithm~\yrcite{anand2021regression}, we still present our algorithm here because it leads to an improved learning-augmented algorithm, as will be presented in Section~\ref{sec:det-pred}.

For simplicity of presentation, we will describe the algorithm's execution as if it never terminates, or in other words, there always comes another new skiing day; however, the actual algorithm is to terminate as soon as it learns that the last day has been reached.
For any $j \geq 1$, let $b(j)$ be a solution (or its cost) covering the most number of days among those whose cost does not exceed $j$, i.e., $b(j) := \opt(t^\star)$ where $t^\star := \max \{ t \in \mathbb{Z}_+ \cup \{\infty\} \mid \opt(t) \leq j\}$. That is, $b(j)$ is the ``best'' thing to do within a budget of $j$.

\subsection{Competitive Algorithm}\label{sec:det-comp}
The algorithm runs in several \emph{iterations}. For each iteration~$i$ (for $i = 1, 2, \cdots$), let $\sol_i$ be the total (starting from the very first iteration) cost of our solution at the end of the $i$-th iteration. In the first iteration, we append $\opt(1)$ to our solution. We thus have $\sol_1 := \opt(1)$. Let  $\tau_1 := 1$. In each later iteration $i \geq 2$, we append $b(\sol_{i - 1})$ to our solution. Let $\tau_i$ be the number of days \emph{newly} covered in iteration $i$, or in other words, the number of days covered by $b(\sol_{i - 1})$.  Remark that, if $b(\sol_{i - 1})$ includes a buy option (i.e., an option $j$ with $d_j=\infty$), $\tau_i$ becomes $\infty$ and no further iterations exist.

We note that the difference from Anand et al.'s algorithm~\yrcite{anand2021regression} is the fact that we append $b(\sol_{i - 1})$ instead of $b(2 \opt(\tau_{i - 1}))$ at each iteration $i$, as is inspired by Zhang et al.~\yrcite{zhang2011ski}.

\begin{thm} \label{thm:det-comp}
This algorithm is 4-competitive.
\end{thm}
The proof of the theorem is deferred to Appendix~\ref{app:det}.

\subsection{Learning-Augmented Algorithm}\label{sec:det-pred}
In this subsection, we describe our deterministic learning-augmented algorithm. We are given the prediction $\Tpred$ on $T$ and the level of confidence $\lambda \in [0, 1]$.
The algorithm consists of (at most) three \emph{phases} as follows. 

\paragraph{First Ignore Phase} We enter this phase at the very beginning if $\opt(1) \leq \lambda \opt(\Tpred)$. Otherwise, we directly enter the respect phase. In this phase, we simply run the previous 4-competitive algorithm. Let $\istar$ be the first iteration where the total cost incurred by the competitive algorithm exceeds $\lambda \opt(\Tpred)$, i.e.,
\begin{equation} \label{eq:pred-firststop}
\sol_{\istar - 1} \leq \lambda \opt(\Tpred) < \sol_{\istar}.
\end{equation}
Note that there always exists such $\istar\geq 2$ since we enter this phase only if $\opt(1) \leq \lambda \opt(\Tpred)$. After processing the iteration~$\istar$, we move on to the respect phase or the second ignore phase depending on $\sol_{\istar}$. If $\sol_{\istar} \leq \opt(\Tpred)$, we enter the respect phase; otherwise, we enter the second ignore phase.

\paragraph{Respect Phase}
Intuitively, this phase is where the algorithm respects the prediction. In this phase, we append $\opt(\Tpred)$ and then move on to the second ignore phase.

\paragraph{Second Ignore Phase}
In this phase, we run the 4-competitive algorithm with a slight modification as follows.

First, let $\sol'_0$ be the total price incurred so far.
Therefore, if the preceding phase was the respect phase, we have $\sol'_0 := \sol_{\istar} + \opt(\Tpred)$ (or $\sol'_0 := \opt(\Tpred)$ if the first ignore phase was skipped). On the other hand, if the preceding phase was the first ignore phase, we have $\sol'_0 := \sol_{\istar}$.

We then choose $\tau'_0$ 
so that it becomes a lower bound on the number of days covered so far.
If the preceding phase was the respect phase, we choose $\tau'_0 := \Tpred$. If the immediately preceding phase was the first ignore phase, we choose $\tau'_0 := \tau_{\istar}$.

Now, for each iteration $i \geq 1$, we append $b(\sol'_{i - 1})$ into our solution and let $\tau'_i$ be the number of days covered by $b(\sol'_{i - 1})$.
This is the end of the algorithm description.

Following is the main theorem for this learning-augmented algorithm. We defer the proof to Appendix~\ref{app:det}.
\begin{thm} \label{thm:det-pred}
The algorithm is a deterministic $\max\{1 + 2\lambda, 4\lambda\}$-consistent $\left( 2 + \frac{2}{\lambda} \right)$-robust algorithm.
\end{thm}

\section{Randomized Algorithms} \label{sec:rand}
In this section, we give randomized algorithms for the multi-option ski rental problem. We present our $e$-competitive algorithm first and then our learning-augmented algorithm which is, for any given $\lambda \in [0, 1]$, $\con(\lambda)$-consistent and $(e^{\lambda} / \lambda)$-robust, where
\[
\con(\lambda) := \begin{cases}
1 + \lambda, & \text{if } \lambda < \frac{1}{e}, \\
(e + 1) \lambda - \ln \lambda - 1, & \text{if } \lambda \geq \frac{1}{e}.
\end{cases}
\]

As in the previous section, we will describe the algorithm's execution as if it never terminates. Recall also that, for any $j \geq 1$, $b(j)$ denotes a solution (or its cost) covering the most number of days among those whose cost does not exceed $j$. If $j < \opt(1)$, let $b(j) := \emptyset$ be an empty solution.

\subsection{Competitive Algorithm} \label{sec:rand-comp}
Let $\sol$ be the solution we maintain, initially $\sol := \emptyset$. At the very beginning, we sample $\alpha \in [1, e)$ from a distribution whose probability density function is $f(\alpha) := 1/\alpha$. The algorithm then runs in \emph{phases}. In phase~$i$ (for $i = 0, 1, \cdots$), we append $b(\alpha e^i)$ to $\sol$. 

\begin{thm} \label{thm:rand-comp}
The given algorithm is a randomized $e$-competitive algorithm.
\end{thm}
The proof of this theorem is deferred to Appendix~\ref{app:rand-comp}.

\subsection{Learning-Augmented Algorithm} \label{sec:rand-pred}
Now we present our randomized learning-augmented algorithm. Recall that we are given a prediction $\Tpred$ and the level of confidence $\lambda \in [0, 1]$.

\paragraph{Assumptions}
We need several assumptions to describe the algorithm.
First, let us assume that $\opt(\Tpred) = e^k$ for some integer $k$.
This assumption is without loss of generality since, if we have $e^{k - 1} < \opt(\Tpred) < e^k$, we may multiply the cost of every option by $\frac{e^k}{\opt(\Tpred)}$. We also assume $\lambda \in (0, 1)$; we will consider the cases for $\lambda \in \{0, 1\}$ at the end of this section.
Finally, let us assume that $\lambda \opt(\Tpred) \geq e$. Observe that we can easily insist this assumption by multiplying the cost of every option by an appropriate power of $e$.

In what follows, we may write $\lambda := e^{-q-r}$ for some $q \in \{0, 1, 2, \cdots, k - 1\}$ and $r \in (0, 1]$. Here we note the range of $r$: it is strictly positive. For example, if $\lambda = e^{-i}$ for some integer $i$, we regard this $\lambda$ as of $q = i - 1$ and $r = 1$.

\paragraph{Algorithm Description}
Let $\A$ be the algorithm presented in Section~\ref{sec:rand-comp}. We run $\A$ and run the same phases as $\A$. On each phase~$i$, we append the same solution as $\A$ with one following exception: If $\alpha  e^i \in [\lambda \opt(\Tpred), \opt(\Tpred)) = [e^{k-q-r}, e^k)$, we append $\opt(\Tpred)$ instead of $b(\alpha  e^i)$ at this phase. If $\opt(\Tpred)$ has already been appended in a previous phase, we simply do nothing instead of appending it once more.

\begin{thm} \label{thm:rand-pred}
For $\lambda \in (0, 1)$, this algorithm is $\con(\lambda)$-consistent and $(e^\lambda/\lambda)$-robust where $\con$ is defined as follows:
\[
\con(\lambda) := \begin{cases}
1 + \lambda, & \text{if } \lambda < \frac{1}{e}, \\
(e + 1) \lambda - \ln \lambda - 1, & \text{if } \lambda \geq \frac{1}{e}.
\end{cases}
\]
\end{thm}

\subparagraph{Consistency Analysis}
Let us begin with showing that our algorithm is $\con(\lambda)$-consistent. Let $\sol$ be the total cost that our algorithm incurs until the end.
For each phase~$i = 0, 1, \cdots, k-q-2$, the algorithm appends $b(\alpha e^i)$ as the same as $\A$. Therefore, up to phase~$(k-q-2)$, our algorithm also incurs in expectation at most
\begin{equation} \label{eq:ranpred-con-pre}
\int_1^e \left( \sum_{i = 0}^{k-q-2} \alpha e^i \right) f(\alpha) d\alpha \leq e^{k-q-1}.
\end{equation}

Let us assume for now that $\lambda < e^{-1}$, i.e., $q \geq 1$. Suppose the algorithm enters phase~$(k-q-1)$. If $\alpha e^{k-q-1} \geq e^{k-q-r}$ (or simply $\alpha \geq e^{1-r}$), the algorithm appends $\opt(\Tpred)$ and terminates then. Otherwise if $\alpha < e^{1 - r}$, the algorithm appends $b(\alpha e^{k - q- 1})$ and may proceed to the next phase. In phase~$(k - q)$, the algorithm appends $\opt(\Tpred)$ and terminates since $\alpha e^{k - q} \in [e^{k-q-r}, e^k)$. Together with Equation~\eqref{eq:ranpred-con-pre}, the total expected cost of our algorithm can be bounded by
\begin{align}
\E[\sol] & \leq e^{k-q-1} + \int_1^{e^{1-r}} \alpha e^{k-q-1} f(\alpha) d \alpha + e^k \nonumber \\
& \leq e^{k - q - r} + e^k   \label{eq:ranpred-cons-q1} \\
& = (1 + \lambda) \cdot \opt(\Tpred). \nonumber
\end{align}

We now turn to the case where $\lambda \leq e^{-1}$, i.e., $q = 0$. Here we have one distinction from the previous case; when the algorithm enters phase~$(k - 1)$, if $\alpha < e^{1-r}$, the algorithm not only appends $b(\alpha  e^{k - 1})$, but may also proceed to phase~$k$ and append $b(\alpha  e^k)$ since $\alpha e^k \geq \opt(\Tpred)$. Therefore, again by Equation~\eqref{eq:ranpred-con-pre},
\begin{align}
\E[\sol] & \leq e^{k-1} + \int_1^{e^{1 - r}} (\alpha e^{k - 1} + \alpha e^k) f(\alpha) d\alpha \nonumber \\
& \qquad + \int_{e^{1 - r}}^e e^k f(\alpha) d\alpha \nonumber \\
& \leq e^{k + 1 -r} + e^{k - r} + e^k(r - 1)  \label{eq:ranpred-cons-q0}  \\
& = \left( (e + 1) \lambda - \ln \lambda - 1 \right)  \opt(\Tpred)  \nonumber
\end{align}
where the last equality holds since $\lambda = e^{-r}$.

\subparagraph{Robustness Analysis}
Let us now show that our algorithm is $(e^\lambda/\lambda)$-robust. Here we break down into several cases as follows depending on $\opt(T)$.

\subparagraph{Case 1. $\opt(T) < e^{k - q - 2}$.}
Note that, in this case, the algorithm executes the same as $\A$, yielding $\E[\sol] \leq e \opt(T)$ by Theorem~\ref{thm:rand-comp}.

\subparagraph{Case 2. $e^{k - q - 2} \leq \opt(T) < e^{k - q -1}$.}
Let $\opt(T) = \beta e^{k - q - 2}$ for some $\beta \in [1, e)$. Observe that, up to phase~$(k - q - 2)$, the algorithm appends the same solution as $\A$.
If $\alpha \geq \beta$ in $\A$, the algorithm terminates at phase~$(k - q - 2)$. Otherwise, the algorithm may proceed to phase~$(k-q-1)$. 

Suppose $\beta < e^{1 - r}$. Observe that, if the algorithm enters phase~$(k-q-1)$, we have $\alpha e^{k-q-1} < \lambda \opt(\Tpred)$, implying that the algorithm executes the same as $\A$. We thus have $\E[\sol] \leq e \opt(T)$ again by Theorem~\ref{thm:rand-comp}.

Let us now assume that $\beta \geq e^{1 - r}$. Observe that, no matter whether $q \geq 1$ or $q = 0$, if $\alpha < e^{1 - r}$, the algorithm appends $b(\alpha e^{k - q - 1})$ by following $\A$ and terminates; otherwise if $e^{1 - r} \leq \alpha < \beta$, the algorithm appends $\opt(\Tpred)$ and terminates. We therefore have
\begin{align*}
\E[\sol] & \leq \int_1^e \left( \sum_{i = 0}^{k - q -2} \alpha e^i \right) f(\alpha) d\alpha \\
& \qquad + \int_1^{e^{1 - r}} \alpha  e^{k - q - 1} f(\alpha) d\alpha \\
& \qquad + \int_{e^{1-r}}^\beta e^k f(\alpha) d\alpha \\
& = \left( e^{k - q - 1} - 1 \right) + (e^{1 - r} - 1) e^{k - q - 1} \\
& \qquad + (\ln \beta + r - 1)e^k \\
& \leq e^{k-q-r} + (\ln \beta + r - 1)e^k.
\end{align*}
Now we can upper bound the robustness ratio for this case as follows:
\begin{align*}
\frac{\E[\sol]}{\opt(T)} & \leq \frac{e^{k-q-r} + (\ln \beta + r - 1) e^k}{\beta e^{k - q- 2}} \\
 & = \frac{e^{2 - r}}{\beta} + \frac{(\ln \beta + r - 1) e^{q + 2}}{\beta} \\
 & = \frac{e^{2 - r}}{\beta} + \frac{(\ln \beta + r - 1) e^{2 - r}}{\lambda \beta}
\end{align*}
where the last equality comes from the definition of $\lambda = e^{-q-r}$. The following claim completes the proof for this case.
\begin{claim} \label{claim:rand1}
We have $h(\beta) := \frac{e^{2 - r}}{\beta} + \frac{(\ln \beta + r - 1) e^{2 - r}}{\lambda \beta} \leq \frac{e^\lambda}{\lambda}$ for every $\beta > 0$.
\end{claim}
The claim can be shown by a simple calculus, which is deferred to Appendix~\ref{app:rand-pred}.

\subparagraph{Case 3. $e^{k - q - 1} \leq \opt(T) < \lambda \opt(\Tpred) = e^{k - q - r}$.}
Let $\opt(T) = \beta e^{k - q - 1}$ for some $\beta \in [1, e^{1 - r})$. Let us assume for now that $q \geq 1$. Note that the execution of our algorithm can be described as the following steps.
\begin{enumerate}[(a)]
\item For each phase $i = 0, 1,\cdots, k - q - 2$, the algorithm appends the same solution as $\A$.
\item On phase~$(k - q - 1)$, if $\alpha < e^{1 - r}$, the algorithm follows $\A$ by appending $b(\alpha e^{k - q- 1})$. Otherwise, the algorithm appends $\opt(\Tpred)$. If $\alpha \geq \beta$, the algorithm immediately terminates then.
\item \label{step3:3} If $\alpha < \beta < e^{1 - r}$, the algorithm enters phase~$(k-q)$ and appends $\opt(\Tpred)$ since $\alpha e^{k - q} \in [e^{k-q-r}, e^k)$ for $q \geq 1$. The algorithm then terminates.
\end{enumerate}
From this description, we can bound the total expected cost incurred by the algorithm as follows:
\begin{align*}
\E[\sol] & \leq  \int_1^e \sum_{i = 0}^{k - q -2} \alpha e^i f(\alpha) d\alpha + \int_1^{e^{1 - r}} \alpha e^{k - q - 1} f(\alpha) d\alpha \\
& \qquad+ \int_{e^{1-r}}^e e^k f(\alpha) d\alpha + \int_{1}^\beta e^k f(\alpha) d\alpha \\
& \leq e^{k - q - r} + e^k \cdot (\ln \beta + r),
\end{align*}
yielding that the robustness ratio for this case can be bounded from above by
\[
\frac{e^{k - q -r} + e^k \cdot (\ln \beta + r)}{\beta \cdot e^{k - q- 1}} = \frac{e^{1-r}}{\beta} + \frac{e^{1 - r}(\ln \beta + r)}{\lambda \beta}
\]
where the equality holds since $\lambda = e^{-q-r}$ by definition. We claim that the right-hand side can be further bounded by $\frac{e^\lambda}{\lambda}$, completing the proof for this case. The proof for this claim can be found in Appendix~\ref{app:rand-pred}.
\begin{claim} \label{claim:rand2}
We have $h(\beta) := \frac{e^{1 - r}}{\beta} + \frac{e^{1 - r}(\ln \beta + r)}{\lambda \beta} \leq \frac{e^\lambda}{\lambda}$ for every $\beta > 0$.
\end{claim}

Now we turn to the case where $q = 0$, i.e., $\opt(T) = \beta \cdot e^{k - 1}$ and $\lambda = e^{-r}$. Observe that there exists one difference on the execution from that of the case where $q \geq 1$; in Step \ref{step3:3}, the algorithm appends $b(\alpha \cdot e^k)$ instead of $\opt(\Tpred)$ since the algorithm now enters phase~$k$ and hence $\alpha \cdot e^k \not\in [e^{k-q-r}, e^k)$. Note that the total expected price for this case can be bounded by
\begin{align*}
\E[\sol] & \leq  \int_1^e \sum_{i = 0}^{k  -2} \alpha e^i f(\alpha) d\alpha + \int_1^{e^{1 - r}} \alpha e^{k - 1} f(\alpha) d\alpha \\
& \qquad + \int_{e^{1-r}}^e e^k f(\alpha) d\alpha + \int_{1}^\beta \alpha \cdot e^k f(\alpha) d\alpha \\
& \leq e^{k - r} + e^k \cdot (\beta + r - 1),
\end{align*}
resulting in the following upper bound for the robustness ratio for this case:
\[
\frac{e^{k - r} + e^k \cdot (\beta + r - 1)}{\beta \cdot e^{k - 1}} = e \cdot \left( 1 + \frac{\lambda + r - 1}{\beta} \right) \leq e \cdot (\lambda + r),
\]
where the inequality can be derived from the fact that $\beta \geq 1$. The following claim completes the proof for this case. Recall that $\lambda = e^{-r}$, and hence $r = - \ln \lambda$, for this case.
\begin{claim} \label{claim:rand3}
We have $e (x - \ln x) \leq \frac{e^x}{x}$ for every $x > 0$.
\end{claim}
We defer the proof of this claim to Appendix~\ref{app:rand-pred}.

It remains to show the cases where $\opt(T) \geq \lambda \opt(\Tpred)$. We again defer the proof for these remaining cases to Appendix~\ref{app:rand-pred}.

\paragraph{Final Remark}
In our analysis, we assume that $\lambda \in (0, 1)$. However, we can see that Theorem~\ref{thm:rand-pred} still holds when $\lambda = 0$ or $\lambda = 1$. In fact, if $\lambda = 0$, this algorithm may correspond to appending $\opt(\Tpred)$ at the very beginning. Observe that this is $1$-consistent, yet $\infty$-robust. On the other hand, if $\lambda = 1$, the algorithm is exactly the same as $\A$ itself, implying that the algorithm is $e$-consistent and $e$-robust.

\section{Lower Bounds} \label{sec:lbound}
In this section, we present lower bounds for the multi-option ski rental problem.

\subsection{Auxiliary Problem} \label{sec:lbound-button}
We define an auxiliary problem which we call the \emph{button problem}. In this problem, we are given an ordered list of $m$ buttons where some buttons are designated as \emph{targets}. We know that the ``targetness'' of the buttons is monotone, i.e., there exists some $J \leq m$ such that buttons~$1$ to~$(J - 1)$ are not targets, but buttons~$J$ to $m$ are all targets. However, we do not know in advance the first target button (i.e., button $J$). To sense whether button $j$ is a target, we must click it paying $b_j \in \mathbb{Q}_+$ as the price; we then learn whether this button is a target or not. The prices of the buttons are all given at the beginning. We also assume that the prices are nondecreasing, i.e., $b_1 \leq b_2 \leq \cdots \leq b_m$. The algorithm clicks buttons until it clicks a target button, and the natural objective is to minimize the total price.

We say that an algorithm for the button problem is $\gamma$-competitive if $\E[\sol] \leq \gamma \cdot b_J$, where $\sol$ is the total price that the algorithm incurs until it clicks a target button. In the learning-augmented version of the problem, the algorithm is given a prediction $\Jpred$ on the first target button~$J$. We say that an algorithm for the button problem is $\chi$-consistent if $\E[\sol] \leq \chi \cdot b_{\Jpred}$ when the prediction is accurate (i.e., $\Jpred=J$), and the algorithm is $\rho$-robust if $\E[\sol] \leq \rho \cdot b_J$ for all $J$ regardless how accurate the prediction $\Jpred$ is.

Lemma~\ref{lem:stb} states that
a lower bound for the button problem immediately extends to give (almost) the same lower bound  for the multi-option ski rental problem.
\begin{lem} \label{lem:stb}
Suppose there exists a randomized $\chi$-consistent $\rho$-robust algorithm for the multi-option ski rental problem with $1\leq \chi \leq \rho$. Then, for all constant $\varepsilon \in (0, 1)$, there exists a randomized $(\chi + \varepsilon)$-consistent $(\rho + \varepsilon)$-robust algorithm for the button problem.
\end{lem}
Let us present the reduction algorithm from the multi-option ski rental problem to the button problem.
Let $\mathcal{A}$ be the $\chi$-consistent $\rho$-robust algorithm for the multi-option ski rental problem.
Without loss of generality, we can assume that the prices $\{b_j\}_{j=1, \cdots, m}$ are all integers; otherwise, we enforce this assumption by simply multiplying the prices by a common denominator. We thus have $b_1 \geq 1$.

Consider the following algorithm for the button problem. Let $C := \ceil{\rho/\varepsilon} \cdot b_m$. Note that $C\geq 2$. The algorithm constructs an instance for the ski rental problem as follows. Let the number of rental options $n$ be $b_m$; set $(d_i,c_i) := (C^i,i)$ for $i = 1, \cdots, n$, and $\Tpred := C^{b_{\Jpred}}$.  The algorithm internally runs $\mathcal{A}$ on this constructed instance. Whenever $\mathcal{A}$ chooses an option, say, option~$i$, we click the \emph{last} button whose cost does not exceed $c_i = i$. (If there does not exist such a button, we do nothing.) If the clicked button turns out to be a target button, we report to $\mathcal{A}$ that the last skiing day has been reached and terminate the whole algorithm. 

However, we might never be able to click a target button with only this procedure since $\mathcal{A}$ may choose only ``cheap'' options. In order to ensure that the algorithm always terminates, we add the following condition. Once the total cost incurred by $\mathcal{A}$ so far becomes at least $C$, we click the last button~$m$ and terminate. We say the algorithm is \emph{forced} to terminate in this case. Let $\sol$ be the cost incurred by the constructed algorithm. Note that $\sol$ is no greater than the cost incurred by $\mathcal{A}$ unless the constructed algorithm is forced to terminate. This is the end of the reduction from the multi-option ski rental problem to the button problem.

The proof of Lemma~\ref{lem:stb} then follows from the next lemma. We defer its proof to Appendix~\ref{app:lbound-button}
\begin{lem} \label{lem:stbproof}
This reduction algorithm is a randomized $(\chi + \varepsilon)$-consistent $(\rho + \varepsilon)$-robust algorithm for the button problem.
\end{lem} 
\subsection{Competitive Ratio of Randomized Algorithms} \label{sec:lbound-rand}
For randomized algorithms, we obtain the following lower bound on the competitive ratio. Together with Theorem~\ref{thm:rand-comp}, this shows that the algorithm presented in Section~\ref{sec:rand-comp} is indeed the best possible.
\begin{thm}\label{thm:lbound-rand}
For all constant $\varepsilon > 0$, no randomized algorithm can achieve a competitive ratio of $e - \varepsilon$.
\end{thm}
Here we give a proof sketch of this theorem. For any instance $\{b_j\}_{j = 1, \cdots, m}$, we can formulate the following LP whose value constitutes a lower bound on the competitive ratio for any randomized algorithm for the button problem.
\begin{align*}
\text{min } & \gamma \\
\text{s.t. } & \textstyle \sum_{j = 1}^m x_j = 1, \\
& \textstyle \sum_{j = t + 1}^{m} y_{t, j} = x_t + \sum_{j = 1}^{t - 1} y_{j, t}, \\
& \qquad \forall t = 1, \cdots, m - 1 \\
& \textstyle \sum_{j = 1}^m b_j \cdot \left(x_j + \sum_{t = 1}^{\min(J, j) - 1} y_{t, j} \right) \leq \gamma \cdot b_J, \\
& \qquad \forall J = 1, \cdots, m, \\
& x_j \geq 0, \quad \forall j = 1, \cdots, m, \\
& y_{t, j} \geq 0, \quad \forall t = 1, \cdots, m - 1, \forall j = t + 1, \cdots, m. \\
\end{align*}
Next, we carefully construct a family of instances and obtain the dual of this LP with respect to this family. We then analytically identify a dual feasible solution whose value converges to $(e - \varepsilon)$. We can then complete the proof of Theorem~\ref{thm:lbound-rand} by the weak duality. The full proof can be found in Appendix~\ref{app:lbound-rand}.
\subsection{Trade-off Between Consistency and Robustness of Deterministic Algorithms} \label{sec:lbound-det}
For deterministic algorithms, we consider a variant of the regular button problem, called the $(K, \beta)$-continuum button problem, where the buttons are given as a continuum on $[1,m]$ with a price function $b:[1,m]\to\mathbb{R}_+$ satisfying that $b$ is a nondecreasing $K$-Lipschitz continuous function and $b(1)=\beta>0$. The following lemma justifies that we can instead consider this variant to obtain a trade-off between consistency and robustness.
\begin{lem} \label{lem:lbound-conti}
Let $f:(0,1)\to\mathbb{R}_+$ be a continuous function. Suppose that, for all $\varepsilon>0$ and $\lambda \in (0,1)$, the robustness of any deterministic $(1+\lambda)$-consistent algorithm for the $(K,\beta)$-continuum button problem must be strictly greater than $f(\lambda)-\varepsilon$. Then, for all $\varepsilon>0$ and $\lambda \in (0,1)$, the robustness of any deterministic $(1+\lambda)$-consistent algorithm for the regular button problem must also be strictly greater than $f(\lambda)-\varepsilon$.
\end{lem}

We first show that no deterministic algorithm for the continuum button problem can achieve a competitive ratio better than $4$; the proof of this statement can be found in Appendix~\ref{app:lbound-det-comp}. Indeed, Zhang et al.~\yrcite{zhang2011ski} provided the same lower bound on the competitive ratio of deterministic algorithms, but we still present our proof since it is useful in obtaining the following trade-off between consistency and robustness  for deterministic learning-augmented algorithms.
\begin{thm}\label{thm:lbound-detpred}
For all constants $\lambda \in (0, 1)$ and $\varepsilon > 0$, the robustness of any deterministic $(1 + \lambda)$-consistent algorithm must be greater than $2 + \lambda + 1/\lambda - \varepsilon$.
\end{thm}
The full proof of this theorem is deferred to Appendix~\ref{app:lbound-det-pred}.

\newcommand{\anand}{\textsf{Anand et al.}}
\newcommand{\ourdet}{\textsf{Our-Det}}
\newcommand{\ourrand}{\textsf{Our-Rand}}

\begin{figure*}[ht]
\vskip 0.2in
\begin{center}
\centerline{\includegraphics[width=0.93\textwidth]{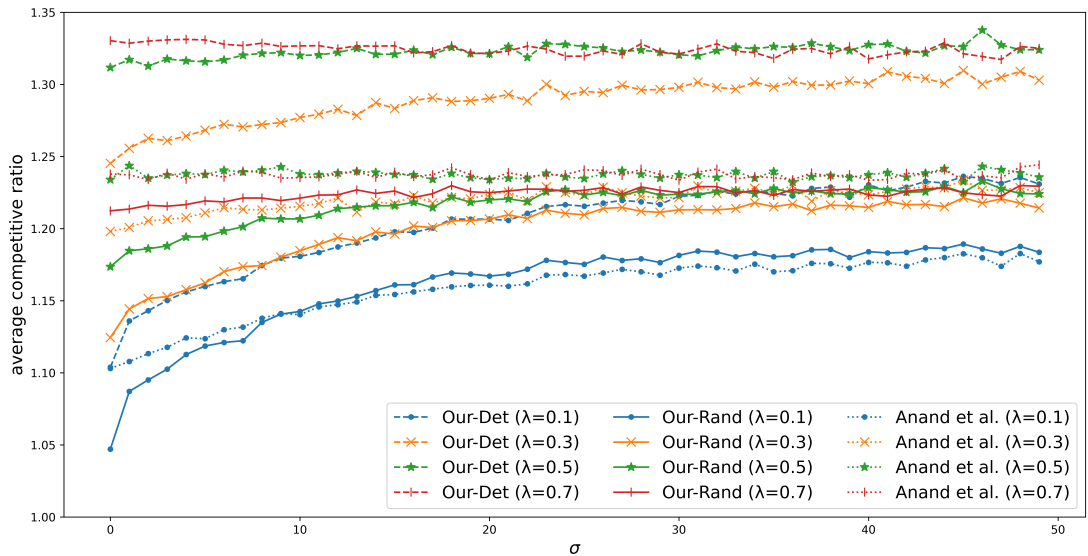}}
\caption{Average competitive ratio achieved by \ourdet{} (dashed line), \ourrand{} (solid line), and \anand{} (dotted line) are shown as a function of $\sigma$. Different colors/markers correspond to different values of the trade-off parameter $\lambda$.}
\label{fig:exp}
\end{center}
\vskip -0.2in
\end{figure*}

\section{Computational Evaluation}\label{sec:exp}

\paragraph{Experiment Setup}
In this section, we computationally 
measure how the average competitive ratios of three learning-augmented algorithms---%
our deterministic  algorithm from Section~\ref{sec:det-pred} (\ourdet{}),
our randomized  algorithm from Section~\ref{sec:rand-pred} (\ourrand{}),
and Anand et al.'s  algorithm (\anand{})%
---behave as the prediction $\hat T$ drifts away from $T$.
For the rent-or-buy (i.e., two-option) ski rental problem, Kumar et al.~\yrcite{kumar2018improving} measured the competitive ratio of their algorithm
by adding to $T$ a Gaussian error in order to obtain prediction $\hat T$. They adjust the standard deviation of the Gaussian distribution to let $\hat T$ drift away from $T$, and conduct experiments for different values of the trade-off parameter $\lambda$.
Our experiments are organized in a similar way: we measure each combination of $\lambda \in \{0.1, 0.3, 0.5, 0.7\}$ and $\sigma \in \{0, 1, \cdots, 50\}$ for all three algorithms; the average competitive ratio of each combination is plotted in Figure~\ref{fig:exp}.

For a fixed $(\lambda, \sigma)$, we generate $10,000$ independent random instances. Unlike Kumar et al.~\yrcite{kumar2018improving}, our experiments need multiple renting options. The rental period and cost of each option is randomly generated; the number of skiing days $T$ is randomly chosen with taking the maximum range of rental period $d^{max}$ into consideration; the prediction error $\eta$ is sampled from $N(0,\sigma)$, with the appropriate clipping and rounding operations\footnote{ The number of renting options is set as $n := 15$ We sample $n$ numbers from $\{1, 2, \cdots, d^{max}:=50\}$ uniformly at random without replacements; let $d_1, \cdots, d_n$ be the sampled numbers, ordered so that $d_1 < \cdots < d_n$.  We next sample $n$ values from $(0, 1)$ uniformly at random independently; let $r_1, \cdots, r_n$ be the sampled values such that $r_1 \le \cdots \le r_n$. We could have just define $c_i:=r_i\cdot d_i$ for all $i$ but this can cause $c_j > c_{j-1}$ for some $j$ which makes option $1$ to option $j-1$ useless. In order to prevent this situation, we inductively construct $c_1, \cdots, c_n$ as follows. Let $c_1 := r_1 \cdot d_1$. For $i = 2, \cdots, n$, if $r_i \cdot d_i < c_{i - 1}$, we multiply $r_i, \cdots, r_n$ by $\frac{c_{i - 1}}{r_i \cdot d_i}$ and add a small constant $10^{-4}$; otherwise, we do nothing. We then set $c_i := r_i \cdot d_i$. This ensures that both $\{d_i\}_{i = 1, \cdots, n}$ and $\{c_i\}_{i = 1, \cdots, n}$ are nondecreasing. Let $\{(d_i, c_i)\}_{i = 1, \cdots, n}$ be the set of renting options. The number of skiing days $T$ is sampled from $\{1, \cdots, 10\cdot d^{max}\}$ uniformly at random. Lastly, we sample error $\eta \sim N(0, \sigma)$ and let $\Tpred := \max\{\lfloor T + \eta\rceil ,1\}$ be the prediction on $T$.}.

\paragraph{Discussion}
As can be seen from Figure~\ref{fig:exp}, \ourrand{} generally outperforms the other two deterministic algorithms. In particular, when $\sigma$ is small, \ourrand{} performs much better than the other two. When $\lambda=0.7$, the plotted lines start to look more ``flat'', showing that the behavior of all three algorithms are less affected by the prediction $\Tpred$. On the other hand, the impact of good predictions was much more dramatic when $\lambda=0.1$. An interesting observation is that \ourdet{} performs slightly worse than \anand{} We believe that this is because \anand{} tends to be more aggressive in using longer rental options compared to  \ourdet{}.

\bibliographystyle{abbrvnat}
\bibliography{lit.bib}

\appendix
\section{Deferred Proofs in Section~\ref{sec:det}} \label{app:det}
To prove Theorems~\ref{thm:det-comp} and~\ref{thm:det-pred}, we need the following observation for the deterministic competitive algorithm presented in Section~\ref{sec:det-comp}.
\begin{obs} \label{obs:det-double}
For each iteration $i \geq 2$, we have $\opt(\tau_i)\leq \sol_{i-1} \leq \opt(\tau_i + 1)$ and $\sol_i \leq 2 \sol_{i - 1}$.
\end{obs}
\begin{proof}
The first series of inequalities follows from the construction. (Note that the only reason why the second inequality is not strict is due to the possibility that $\tau_i=\infty$.) The second inequality follows from the definition of $b$ and the fact that $\sol_i = \sol_{i - 1} + b(\sol_{i - 1})$.
\end{proof}

We are now ready to prove the two main theorems.
\begin{thm}[Theorem~\ref{thm:det-comp} restated] 
This algorithm is 4-competitive.
\end{thm}
\begin{proof}
If $T = \tau_1 = 1$, the algorithm runs optimally. If $\tau_1 + 1 \leq T \leq \tau_2$, note that $\sol_2 \leq 2\sol_1 = 2\opt(1) \leq  2 \opt(T)$. For $i \geq 3$, if $\tau_{i - 1} + 1 \leq T \leq \tau_i$, the algorithm terminates at iteration~$i$ or earlier. We have \[
\sol_i \leq 2 \sol_{i - 1} \leq 4 \sol_{i - 2} \leq 4 \opt(\tau_{i - 1} + 1) \leq 4 \opt(T)
\]
from Observation~\ref{obs:det-double}.
\end{proof}

\begin{thm}[Theorem~\ref{thm:det-pred} restated]
The algorithm is a deterministic $\max\{1 + 2\lambda, 4\lambda\}$-consistent $\left( 2 + \frac{2}{\lambda} \right)$-robust algorithm.
\end{thm}
\begin{proof}
Let us first show the consistency of the algorithm.
If the algorithm starts in the respect phase, it is easy to see that the algorithm is optimal. Thus, we will assume from now on that the algorithm starts in the first ignore phase. If the algorithm moves on to the respect phase, it will terminate at the respect phase. The total cost is exactly $\sol_{\istar} + \opt(\Tpred)$. By Observation~\ref{obs:det-double} and Equation~\eqref{eq:pred-firststop}, this cost is at most $(1 + 2\lambda) \opt(\Tpred)$. If the algorithm terminates during the first ignore phase, imagine we continue the execution of the algorithm until at least the first phase is completed. The total cost in this case is at most $\sol_{\istar}$, and the consistency follows from $\sol_{\istar}\leq 2\lambda \opt(\Tpred)$.

Now consider the case where the algorithm directly enters the second ignore phase after the first ignore phase. This happens when $\sol_{\istar} > \opt(\Tpred)$. Observe that $b(\sol'_0) = b(\sol_{\istar})$ covers at least $\Tpred$ days, implying that the algorithm must terminate after the first iteration, during which we append $b(\sol'_0)$. The total cost therefore can be bounded by
\[
\sol'_1 = 2 \sol_{\istar} \leq 4 \sol_{\istar - 1} \leq 4\lambda \opt(\Tpred)
\]
from Observation~\ref{obs:det-double} and Equation~\eqref{eq:pred-firststop}.

We now turn to proving the robustness. Note that if the algorithm does not enter the respect phase, the algorithm is $4$-robust since the execution of this algorithm is exactly the same as that of our $4$-competitive algorithm from Section~\ref{sec:det-comp}. Therefore, it suffices to consider the case that the algorithm indeed enters the respect phase.

Let us first consider the case where the algorithm starts in the first ignore phase and terminates in the respect phase. From our assumption that the algorithm enters the respect phase, we have $T > \tau_{\istar}$. Observe that
\[
\lambda \opt(\Tpred) < \sol_{\istar} \leq 2\sol_{\istar - 1} \leq 2 \opt(\tau_{\istar} + 1) \leq 2 \opt(T),
\]
where the first inequality follows from Equation~\eqref{eq:pred-firststop} and the
rest from Observation~\ref{obs:det-double} and $T\geq \tau_{\istar} + 1$. This implies
$\opt(\Tpred) \leq \frac{2}{\lambda} \opt(T)$. Therefore, the total cost is at most
$
\sol_{\istar} + \opt(\Tpred) \leq \left( 2 + \frac{2}{\lambda} \right) \opt(T).
$

We now consider the case where the algorithm starts in the first phase, moves on to the respect phase, and terminates in the second ignore phase. 
Suppose that $\tau'_{i - 1} + 1 \leq T \leq \tau'_i$ for some $i \geq 1$. Note that the algorithm terminates at iteration $i$ or earlier then.
Since the algorithm does not terminate in the respect phase, we have $T > \Tpred =\tau'_0$.
If $i = 1$, observe that
\[
\sol'_1 \leq 2 \sol'_0 \leq 4 \opt(\Tpred) \leq 4 \opt(\tau'_{0} + 1) \leq 4 \opt(T)
\]
where the second inequality follows from the fact that the algorithm enters the respect phase only if $\sol_{\istar} \leq \opt(\Tpred)$.
If $i \geq 2$, note that
\begin{equation} \label{eq:det-rob-big}
\sol'_i \leq 4 \sol'_{i - 2} \leq 4 \opt(\tau'_{i - 1} + 1) \leq 4 \opt(T)
\end{equation}
where the first two inequalities follow from Observation~\ref{obs:det-double}.

Finally, consider the case where the algorithm starts in the respect phase. This happens only if $\opt(1) > \lambda \opt(\Tpred)$. If $T \leq \Tpred$, observe that the algorithm terminates in the respect phase and the total cost incurred by the algorithm is at most
\[
\opt(\Tpred) <\frac{1}{\lambda} \opt(1) \leq \frac{1}{\lambda} \opt(T).
\]
If $T > \Tpred$, the algorithm terminates in the second ignore phases. Suppose that $\tau'_{i - 1} + 1 \leq T \leq \tau'_{i}$ for some $i$. The algorithm then terminates at iteration $i$ or earlier. If $i = 1$, we have 
\[
\sol'_1 \leq 2 \sol'_0 = 2 \opt(\Tpred) \leq 2 \opt(\tau'_0 + 1) \leq 2 \opt(T).
\]
If $i \geq 2$, Equation~\eqref{eq:det-rob-big} again holds, completing the proof.
\end{proof}
\section{Deferred Proofs in Section~\ref{sec:rand}}

\subsection{Deferred Proof in Section~\ref{sec:rand-comp}} \label{app:rand-comp}
\begin{thm} [Theorem~\ref{thm:rand-comp} restated]
The algorithm is a randomized $e$-competitive algorithm.
\end{thm}
\begin{proof}
Let $\opt(T) = \beta e^{\istar}$ for some $\istar \in \mathbb{Z}_{\geq 0}$ and $\beta \in [1, e)$. Suppose that the algorithm enters phase~$\istar$; otherwise, the algorithm only incurs less. If $\alpha \geq \beta$, observe that the algorithm terminates at this phase since $b(\alpha e^{\istar})$ covers $T$ days by the definition of $b$ and that $\alpha e^{\istar} \geq \opt(T)$. On the other hand, if we sample $\alpha$ such that $\alpha  < \beta$, the algorithm may enter the next phase~$(\istar + 1)$. Note that the algorithm terminates then since $\alpha e^{\istar + 1} \geq \opt(T)$. Recall that, in each phase~$i$, the algorithm incurs at most $\alpha e^i$. We therefore have
\begin{align*}
\E[\sol] & \leq \int_1^\beta \left( \sum_{i = 0}^{\istar + 1} \alpha e^i \right) f(\alpha) d\alpha + \int_\beta^e  \left( \sum_{i = 0}^{\istar} \alpha e^i \right)  f(\alpha) d\alpha \\
& = \sum_{i = 0}^{\istar} e^i \cdot \int_1^e \alpha f(\alpha) d\alpha + e^{\istar + 1} \cdot \int_1^\beta \alpha f(\alpha) d\alpha \\
& = \frac{e^{\istar + 1} - 1}{e - 1} \cdot (e - 1) + e^{\istar + 1} \cdot (\beta - 1) \\
& = (e^{\istar + 1} - 1) + e^{\istar + 1} \cdot (\beta - 1) \\
& = \beta \cdot e^{\istar + 1} - 1 \leq e \cdot \opt(T).
\end{align*}
\end{proof}

\subsection{Deferred Proof in Section~\ref{sec:rand-pred}} \label{app:rand-pred}
\begin{thm} [Theorem~\ref{thm:rand-pred} restated]
For $\lambda \in (0, 1)$, this algorithm is $\con(\lambda)$-consistent and $(e^\lambda/\lambda)$-robust where $\con$ is defined as follows:
\[
\con(\lambda) := \begin{cases}
1 + \lambda, & \text{if } \lambda < \frac{1}{e}, \\
(e + 1) \lambda - \ln \lambda - 1, & \text{if } \lambda \geq \frac{1}{e}.
\end{cases}
\]
\end{thm}

\paragraph{Robustness Analysis}
We continue to prove the robustness of our randomized learning-augmented algorithm. Let us first prove Claims~\ref{claim:rand1}, \ref{claim:rand2}, and~\ref{claim:rand3}.
\begin{claim} [Claim~\ref{claim:rand1} restated]
We have \[h(\beta) := \frac{e^{2 - r}}{\beta} + \frac{(\ln \beta + r - 1) e^{2 - r}}{\lambda \beta} \leq \frac{e^\lambda}{\lambda}\] for every $\beta > 0$.
\end{claim}
\begin{proof}
Let us obtain the partial derivative of $h(\beta)$ with respect to $\beta$ as follows:
\[
\frac{\partial h(\beta)}{\partial \beta} = \frac{e^{2 - r}}{\lambda} \left( \frac{1 - \ln \beta}{\beta^2} - \frac{\lambda + r - 1}{\beta^2} \right) = \frac{e^{2 - r}}{\lambda \beta^2} \cdot \left( 2 - \lambda - r - \ln \beta \right),
\]
implying that $h(\beta)$ achieves the maximum value of $\frac{e^\lambda}{\lambda}$ at $\beta = e^{2 - \lambda - r}$.
\end{proof}

\begin{claim} [Claim~\ref{claim:rand2} restated]
We have \[h(\beta) := \frac{e^{1 - r}}{\beta} + \frac{e^{1 - r}(\ln \beta + r)}{\lambda \beta} \leq \frac{e^\lambda}{\lambda}\] for every $\beta > 0$.
\end{claim}
\begin{proof}
When we calculate the partial derivative of $h(\beta)$ with respect to $\beta$, we have
\[
\frac{\partial h(\beta)}{\partial \beta} = \frac{e^{1 - r}}{\lambda} \left( \frac{1 - \ln \beta}{\beta^2} - \frac{\lambda + r}{\beta^2} \right) = \frac{e^{1 - r}}{\lambda \beta^2} \cdot \left(1 - \lambda - r - \ln \beta \right),
\]
leading to that $h(\beta)$ achieves the maximum value of $\frac{e^\lambda}{\lambda}$ at $\beta = e^{1 - \lambda - r}$ 
\end{proof}

\begin{claim} [Claim~\ref{claim:rand3} restated]
We have \[e (x - \ln x) \leq \frac{e^x}{x}\] for every $x > 0$.
\end{claim}
\begin{proof}
Let $h(x) := \frac{e^x}{x} - ex + e \ln x$. It suffices to show that $h(x) \geq 0$ for every $x > 0$. Observe that
\[
h'(x) = \frac{e^x (x - 1)}{x^2} - e + \frac{e}{x} = \frac{(x - 1)(e^x - ex)}{x^2},
\]
implying that $h(x)$ has the minimum value of $0$ at $x = 1$.
\end{proof}

It remains to analyze the cases where $\opt(T) \geq e^{k - q - 1}$.
\subparagraph{Case 4. $\lambda \opt(\Tpred) = e^{k - q - r} \leq \opt(T) < \opt(\Tpred) = e^{k}$.}
Remark that, in this case, the algorithm incurs in expectation at most the cost of the case when the prediction is accurate, i.e., $T = \Tpred$.
Therefore, the bounds we obtained in the consistency analysis can be also used in this case. For $q \geq 1$, by Equation~\eqref{eq:ranpred-cons-q1}, we can obtain an upper bound for the robustness ratio for this case as follows:
\[
\frac{\E[\sol]}{\opt(T)} \leq \frac{e^k + e^{k - q- r}}{e^{k - q- r}} = 1 + \frac{1}{\lambda} \leq \frac{e^\lambda}{\lambda}
\]
where the equality comes from the definition of $\lambda = e^{-q-r}$ and the last inequality holds since $1 + x \leq e^x$ for every $x$.

If $q = 0$, by Equation~\eqref{eq:ranpred-cons-q0}, the robustness ratio can be bounded by
\[
\frac{\E[\sol]}{\opt(T)} \leq \frac{e^{k - r} + e^{k + 1 - r} + e^k(r-1)}{e^{k - r}} = 1 + e + \frac{r - 1}{\lambda}
\]
where we derive the equality by the definition of $\lambda = e^{-r}$. Here we claim that the right-hand side can still be bounded by $\frac{e^\lambda}{\lambda}$ as desired for $\lambda$ in this case. Recall that $\lambda \in \left[\frac{1}{e}, 1 \right)$ and $r = - \ln \lambda$.
\begin{claim}
For $x \in \left[\frac{1}{e}, 1 \right]$, we have \[1 + e - \frac{\ln x + 1}{x} \leq \frac{e^x}{x}.\]
\end{claim}
\begin{proof}
Let $h(x) := e^x - (1 + e)x + \ln x + 1$. It suffices to show that $h(x) \geq 0$ for $x \in \left[ \frac{1}{e}, 1 \right]$. Taking the derivative, we can obtain
\[
h'(x) = e^x - (1+e) + \frac{1}{x} = \left( e^x + \frac{1}{x} \right) - (e + 1).
\]
Note that $h'(x)$ is also convex over $x > 0$. Moreover, we have
\[h'\left( \frac{1}{e}\right) = e^{e^{-1}} - 1 > 0, \]
together with $h'(1) = 0$. We can thus conclude that the minimum of $h(x)$ over $\left[ \frac{1}{e}, 1 \right]$ must be either $h\left( \frac{1}{e} \right)$ or $h(1)$. Observe that $h(1) = 0$ and
\[ h\left( \frac{1}{e} \right) = e^{e^{-1}} - \frac{1 + e}{e} + \ln \left( \frac{1}{e} \right) + 1 = e^{e^{-1}} - \left(1 + \frac{1}{e} \right) \geq 0 \]
where the inequality follows from that $e^x \geq 1 + x$ for every $x$.
\end{proof}

\subparagraph{Case 5. $\opt(T) > \opt(\Tpred) = e^k$.}
Let $\opt(T) := \beta \cdot e^{k + p}$ for some $p \in \mathbb{Z}_{\geq 0}$ and $\beta \in [1, e)$. Let us first consider the case where $q \geq 1$. Observe that, until the algorithm reaches phase~$k$, the algorithm incurs in expectation at most $e^k + e^{k - q - r}$ by Equation~\eqref{eq:ranpred-cons-q1}. Note also that, from phase~$k$, we again follow $\A$. In total, we have
\begin{align*}
\E[\sol] & \leq e^k + e^{k - q - r} + \int_1^e \left( \sum_{i = k}^{k+p} \alpha \cdot e^i \right) f(\alpha) d\alpha + \int_1^\beta \alpha \cdot e^{k + p + 1} f(\alpha) d\alpha \\
& = \beta \cdot e^{k + p + 1} + e^{k - q - r},
\end{align*}
implying that the robustness ratio for this case can be bounded by
\[
\frac{\beta \cdot e^{k + p + 1} + e^{k - q - r}}{\beta \cdot e^{k + p}} = e + \frac{\lambda}{\beta e^p} \leq e + \lambda
\]
where the equality follows from the definition of $\lambda = e^{-q-r}$ and the inequality can be derived by the fact that $\beta \geq 1$ and $p \geq 0$. Now the proof can be completed by the following claim. Recall that $\lambda < \frac{1}{e}$ in this case.
\begin{claim}
For $x \in \left(0, \frac{1}{e} \right]$, we have \[ x + e \leq \frac{e^x}{x}. \]
\end{claim}
\begin{proof}
Note that $\frac{e^x}{x}$ is decreasing over $x \in (0, 1]$ while $x + e$ is increasing over $x \in \mathbb{R}$. Direct calculation gives
\[ \frac{e^{e^{-1}}}{e^{-1}} > e + \frac{1}{e} \]
as claimed.
\end{proof}

Finally, let us assume that $q = 0$, i.e., $\lambda = e^{-r}$. Observe that, in this case, the algorithm executes almost identical to $\A$ with only exception that, on phase~$(k - 1)$, the algorithm appends $\opt(\Tpred)$ instead of $b(\alpha \cdot e^{k - 1})$ if $\alpha \geq e^{1-r}$. Therefore, by adjusting the proof of Theorem~\ref{thm:rand-comp} accordingly, we can derive
\begin{align*}
\E[\sol] & \leq \int_1^e \left( \sum_{i = 0}^{k - 2} \alpha \cdot e^i \right) f(\alpha) d\alpha + \int_1^{e^{1-r}} \alpha \cdot e^{k-1} f(\alpha) d\alpha + \int_{e^{1-r}}^e e^k f(\alpha) d\alpha \\
& \quad \quad + \int_1^e \left( \sum_{i = k}^{k + p} \alpha \cdot e^i \right) f(\alpha) d\alpha + \int_1^\beta \alpha \cdot e^{k + p + 1} f(\alpha) d\alpha \\
& \leq e^{k - r} + r \cdot e^k + \beta \cdot e^{k + p + 1} - e^k \\
& = \beta \cdot e^{k + p + 1} + (\lambda + r - 1) \cdot e^k.
\end{align*}
We can thus obtain an upper bound of the robustness ratio for this case as follows:
\[
\frac{\E[\sol]}{\opt(T)} \leq e + \frac{\lambda + r - 1}{\beta e^p} \leq e + \lambda + r - 1
\]
where the last inequality is due to the fact that $\beta \geq 1$ and $p \geq 0$. Note that the next claim completes the proof for this case. Recall that $r = - \ln \lambda$.
\begin{claim}
We have
\[ x - \ln x + e - 1 \leq \frac{e^x}{x} \]
for any $x > 0$. 
\end{claim}
\begin{proof}
Let $h(x) := \frac{e^x}{x} - x + \ln x$. It suffices to show that $h(x) \geq e - 1$ for all $x > 0$. Note that
\[
h'(x) = \frac{e^x (x - 1)}{x^2} - 1 + \frac{1}{x} = \frac{(e^x - x) (x - 1)}{x^2},
\]
implying that the minimum value of $h(x)$ for $x > 0$ is $h(1) = e - 1$.
\end{proof}
\section{Deferred Proofs in Section~\ref{sec:lbound}}

\subsection{Deferred Proof in Section~\ref{sec:lbound-button}} \label{app:lbound-button}
\begin{lem} [Lemma~\ref{lem:stbproof} restated]
This reduction algorithm is a randomized $(\chi + \varepsilon)$-consistent $(\rho + \varepsilon)$-robust algorithm for the button problem.\end{lem}

\begin{proof}
Let us first show that this  algorithm is $(\rho + \varepsilon)$-robust. Let $k$ be the price of the first target button, i.e., $k = b_J$. Now imagine the execution of $\mathcal{A}$ with $T = C^k$. It is easy to see that $\opt(T) = k = b_J$ where $\opt(T)$ is the cost of an optimal solution to cover $T$ days. Observe that, if $k = 1$, the reduction algorithm immediately terminates after $\mathcal{A}$ chooses any option. On the other hand, the first option chosen by $\mathcal{A}$ must be of cost at most $\rho$ in expectation, since $\mathcal{A}$ must be $\rho$-competitive even when $T=1$ (and $\opt(T)=1$). This implies the $\rho$-robustness of the constructed algorithm for the button problem.
Thus, we assume from now on that $k \geq 2$. This implies $n=b_m\geq b_J\geq 2$.
 
We claim that, if $\A$ covers $T$ days only with options~$1$ to $(k - 1)$, the total cost $c(\A)$ incurred by $\A$ is at least $C$. To see this fact, let $a_i$  be the number of times that $\A$ chooses option~$i$, for $i = 1, \cdots, k - 1$. As the output of $\A$ covers $T$ days, we have
\begin{equation*} \label{eq:stb1}
\sum_{i = 1}^{k - 1} C^i \cdot a_i \geq T = C^k,
\end{equation*}
yielding
\[
\sum_{i = 1}^{k - 1} \frac{k - 1}{C^{k - 1 - i}} \cdot a_i \geq C \cdot (k - 1)
.\]
We can now show our claim as follows:
\begin{equation*} \label{eq:stb2}
c(\mathcal{A}) = \sum_{i = 1}^{k - 1} i \cdot a_i \geq \sum_{i = 1}^{k - 1} \frac{k - 1}{C^{k - 1 - i}} \cdot a_i \geq C \cdot (k - 1) \geq C
\end{equation*}
where the first inequality holds since $h(x) := (k - 1)/C^{k - 1 - x}$ is a convex function satisfying $h(1) = (k - 1)/C^{k - 2} \leq 1$ and and $h(k - 1) = k - 1$. Recall that $k \geq 2$ and $C \geq 2$.

The above claim implies that, if $c(\mathcal{A}) < C$, $\mathcal{A}$ chose an option whose cost is at least $k = b_J$. Therefore, the constructed algorithm will not be forced to terminate in this case, and we have
\begin{equation}\label{eq:stb2_5}
\sol \leq c(\mathcal{A}).
\end{equation}
On the other hand, if $c(\mathcal{A})\geq C$, the constructed algorithm is forced to terminate with an additional price of
\begin{equation} \label{eq:stb3}
b_m = n \leq \frac{\varepsilon}{\rho} \cdot C \leq \frac{\varepsilon}{\rho} \cdot c(\mathcal{A})
,\end{equation}yielding\begin{equation}\label{eq:stb3_5}
\sol \leq \left( 1+\frac{\varepsilon}{\rho} \right)c(\mathcal{A})
.\end{equation}
Equations~\eqref{eq:stb2_5} and~\eqref{eq:stb3_5} together implies that $\sol \leq \left(1 + \frac{\varepsilon}{\rho} \right) \cdot c(\mathcal{A})$ holds whether or not $c(\mathcal{A}) < C$. Hence, we have
\begin{equation} \label{eq:stb4}
\E[\sol] \leq \left( 1 + \frac{\varepsilon}{\rho} \right) \cdot \E[c(\mathcal{A})] \leq (\rho + \varepsilon) \cdot b_J,
\end{equation}
where the second inequality is derived from that $\mathcal{A}$ is $\rho$-robust.

To see that the constructed algorithm is $(\chi + \varepsilon)$-consistent, note that $b_m \leq (\varepsilon/\chi) \cdot c(\mathcal{A})$ from Equation~\eqref{eq:stb3} since $\chi \leq \rho$. By adjusting Equation~\eqref{eq:stb4} with the fact that $\mathcal{A}$ is $\chi$-consistent for $\Tpred = C^{b_{\Jpred}}$, we can obtain $\E[\sol] \leq (\chi + \varepsilon) \cdot b_{\Jpred}$ as desired.

Finally, let us remark that the above argument only holds when $\rho$ is finite. Suppose $\mathcal{A}$ is non-robust (i.e., $\rho=\infty$). In this case, we do not need to consider the robustness of the constructed algorithm either. By setting $C := \ceil{\chi/\varepsilon} \cdot b_m$, we can see that the above argument for the consistency still follows.
\end{proof}
\subsection{Deferred Proof in Section~\ref{sec:lbound-rand}} \label{app:lbound-rand}
\begin{thm}[Theorem~\ref{thm:lbound-rand} restated]
For all constant $\varepsilon > 0$, no randomized algorithm can achieve a competitive ratio of $e - \varepsilon$.
\end{thm}
\begin{proof}
By Lemma~\ref{lem:stb}, it suffices to exhibit a family of instances for the button problem that makes any algorithm have a competitive ratio at least $(e - \varepsilon)$ for the given constant $\varepsilon > 0$.

Now we introduce a set of parameters that define an instance. It is important in which order we choose these parameters, but we will discuss this at the end of the proof. For now, let $\delta$ be a sufficiently large number; intuitively speaking, $\delta$ corresponds to the granularity of button prices. Parameter $c$ satisfies $c/\delta\geq- \ln(e - \varepsilon) - \ln \ln \left(e/(e - \varepsilon)\right)$. An interger $m$ is chosen so that $m > c$ and $m/\delta \geq \ln \left( e/\varepsilon \right)$. Consider the following instance of the button problem defined by these parameters: the number of buttons is $m$ and their prices are  $b_j := e^{j/\delta}$ for $j = 1, \cdots, m$.

Fix any algorithm for the button problem. Without loss of generality, we can assume that the indicies of the buttons clicked by the algorithm strictly increases. That is, if the algorithm clicks a button $j$ at some point, it will click button $j'>j$ in the following rounds. Suppose for the moment that only the last button $m$ is a target: i.e, $J=m$. For $j = 1, \cdots, m$, let $x_j$ be the marginal probability that the algorithm clicks button~$j$ in the first round. For $t = 1, \cdots, m - 1$ and $j = t + 1, \cdots, m$, let $y_{t, j}$ be the marginal probability that the algorithm clicks button~$t$ in some round and then clicks button~$j$ in the immediately following round.

Observe that $x_j$ does not depend on $J$: the algorithm chooses the first button without any knowledge on $J$ anyways. Similarly, for all $t<J$, $y_{t, j}$ does not depend on $J$ either. Intuitively speaking, for any $J_1 < J_2$, the ``prefix'' of the execution of the algorithm for $J = J_2$ is the same as that for $J = J_1$ due to the algorithm's lack of knowledge on $J$.

We now write an LP where these $x$'s and $y$'s are variables and the constraints specify the properties that must be satisfied by the algorithm. We can write these constraints assuming $J=m$, since we can retrieve the marginal probabilities for the cases where $J\neq m$ simply by taking a ``prefix''. The value of the following LP is a lower bound on the competitive ratio of the algorithm.
\begin{align*}
\text{minimize } & \gamma \\
\text{subject to } & \sum_{j = 1}^m x_j = 1, \\
& \sum_{j = t + 1}^{m} y_{t, j} = x_t + \sum_{j = 1}^{t - 1} y_{j, t}, & \forall t = 1, \cdots, m - 1 \\
& \sum_{j = 1}^m b_j \cdot \left(x_j + \sum_{t = 1}^{\min(J, j) - 1} y_{t, j} \right) \leq \gamma \cdot b_J, & \forall J = 1, \cdots, m, \\
& x_j \geq 0, & \forall j = 1, \cdots, m, \\
& y_{t, j} \geq 0, & \forall t = 1, \cdots, m - 1, \forall j = t + 1, \cdots, m. \\
\end{align*}
Note that the first constraint must be satisfied because $x$'s must form a probability distribution. Note that both sides of the second constraint is a way of writing the marginal probability that the algorithm clicks button $t$ (recall that we assume $J=m$). Therefore, these equalities must be satisfied. Note that $x_j + \sum_{t = 1}^{\min(J', j) - 1} y_{t, j}$ is the marginal probability that button $j$ is clicked when $J=J'$: we use $\min(J', j) - 1$ to ensure that we take an appropriate prefix. This shows that the third constraints must be satisfied as long as $\gamma$ is no smaller than the true competitive ratio.

In order to compute the worst-case value of this LP, we consider its dual, shown below:
\begin{align*}
\text{maximize } & w \\
\text{subject to } & \sum_{j = 1}^m b_j v_j = 1, \\
& w \leq u_t + b_t \sum_{j = 1}^m v_j, & \forall t = 1, \cdots, m - 1 \\
& w \leq b_m \sum_{j = 1}^m v_j, \label{prog:lbound-rand-dual1}\tag{D1} \\
& u_s - u_t \leq b_t \sum_{j = s + 1}^m v_j, & \forall s = 1, \cdots, m - 2, \forall t = s + 1, \cdots, m - 1, \\
& u_s \leq b_m \sum_{j = s + 1}^m v_j, & \forall s = 1, \cdots, m - 1, \\
& w \in \mathbb{R}, \\
& u_t \in \mathbb{R}, & \forall t = 1, \cdots, m - 1, \\
& v_j \geq 0, & \forall j = 1, \cdots, m. \\
\end{align*}
We want to find a feasible solution to \eqref{prog:lbound-rand-dual1} whose value is close to $e - \varepsilon$. To this end, let us consider the following auxiliary LP.
\begin{align*}
\text{maximize } & w \\
\text{subject to } & w \leq u_t + b_t \sum_{j = 1}^m v_j, & \forall t = 1, \cdots, m, \\
& u_s - u_t \leq b_t \sum_{j = s + 1}^m v_j, & \forall s = 1, \cdots, m - 1, \forall t = s + 1, \cdots, m, \label{prog:lbound-rand-dual2}\tag{D2} \\
& u_m = 0, \\
& w \in \mathbb{R}, \\
& u_t \in \mathbb{R}, & \forall t = 1, \cdots, m, \\
& v_j \geq 0, & \forall j = 1, \cdots, m. \\
\end{align*}
Observe that, once we obtain a feasible solution to \eqref{prog:lbound-rand-dual2}, we can easily construct a feasible solution to \eqref{prog:lbound-rand-dual1} by dividing every variable by $\sum_{j = 1}^m b_j v_j$.

Let us construct a feasible solution to \eqref{prog:lbound-rand-dual2} as follows: for all $j = 1, \cdots, m$,
\begin{align*}
v_j & := \int_{(j-1)/\delta}^{j/\delta} e^{-z} dz, \\
u_j & := \frac{e - \varepsilon}{\delta} \left( m - c - j \right)_+, \textrm{ and}  \\
w & := \min_{t = 1, \cdots, m} \left\{ u_t + e^{t/\delta} \sum_{j = 1}^m v_j \right\},
\end{align*}
where $(\cdot)_+ := \max\{\cdot, 0\}$. Observe that the first set of constraints of \eqref{prog:lbound-rand-dual2} is satisfied by the choice of $w$. It is easy to see that the third and fourth sets are also satisfied. The next lemma shows that the solution satisfies the second set of constraints.
\begin{lem}
If $c/\delta \geq - \ln(e - \varepsilon) - \ln \ln \left(e/(e - \varepsilon)\right) $, we have
\[u_s - u_t \leq b_t \sum_{j = s + 1}^m v_j\]
for any $s < t$.
\end{lem}
\begin{proof}
Let us first consider the case where $s < t \leq m - c$. Observe that
\begin{align*}
u_s - u_t & = (e - \varepsilon) \cdot \frac{t - s}{\delta} \ \textrm{ and } \\
b_t \sum_{j = s + 1}^m v_j & = e^{t/\delta} \int_{s/\delta}^{m/\delta} e^{-z} dz = e^{(t-s)/\delta} - e^{(t-m)/\delta}.
\end{align*}
We will show that
\[
(e - \varepsilon) \cdot \frac{t - s}{\delta} \leq e^{(t -s) / \delta} - e^{-c/\delta} \leq e^{(t-s)/\delta} - e^{(t-m)/\delta}
,\]
where the last inequality follows from $t \leq m - c$, showing the lemma for this case. By substituting $z := (t - s)/\delta$ and rearranging the terms, it suffices to show that, for any $z > 0$,
\[
e^z - (e - \varepsilon) \cdot z \geq e^{-c/\delta}.
\]
By taking the derivative of the left-hand side with respect to $z$, we can easily see that the  left-hand side is minimized when $z = \ln (e - \varepsilon)$. Observe that
\[
e^{-c/\delta} \leq (e - \varepsilon) - (e - \varepsilon) \ln (e - \varepsilon)
\]
due to the condition of the lemma, showing the claim.

For $s < m - c \leq t$, we have
\begin{align*}
u_s - u_t & = (e - \varepsilon) \cdot \frac{m - c- s}{\delta} \ \textrm{ and } \\
b_t \sum_{j = s + 1}^m v_j & = e^{t/\delta} \left( e^{-s/\delta} - e^{-m/\delta} \right).
\end{align*}
Note that, for a fixed $s$, $b_t \sum_{j = s + 1}^m v_j$ is minimized when $t = m - c$ while $u_s - u_t$ is unaffected by $t$. Therefore, it suffices to show that the lemma follows when $t = m - c$. This is already proven by the previous case.

Finally, for $m - c \leq s < t$, we have $u_s - u_t = 0$.
\end{proof}

The following lemma gives a lower bound on $w$.
\begin{lem}
If $m/\delta \geq \ln \left( e/\varepsilon \right)$, we have \[h(t) := u_t + e^{t/\delta} \sum_{j = 1}^m v_j \geq \frac{e - \varepsilon}{\delta} (m - c)\] for any $t$.
\end{lem}
\begin{proof}
Recall that, for $t \geq m - c$, $u_t = 0$. Hence, it suffices to consider $t \leq m - c$. We have
\begin{align*}
h(t) & = \frac{e - \varepsilon}{\delta} (m - c - t) + e^{t/\delta} \int_{0}^{m/\delta} e^{-z} dz \\
& = \frac{e - \varepsilon}{\delta} (m - c - t) + e^{t/\delta} \left( 1 - e^{-m/\delta} \right).
\end{align*}
By taking the partial derivative of $h(t)$ with respect to $t$, we obtain
\[
\frac{\partial h(t)}{\partial t} = - \frac{e - \varepsilon}{\delta} + \frac{e^{t / \delta}}{\delta} \cdot \left( 1 - e^{-m/\delta} \right).
\]
This implies that $h(t)$ is minimized when $t = t^\star$ where $t^\star$ satisfies
\[
e^{t^\star/\delta} = \frac{e - \varepsilon}{1 - e^{-m/\delta}} \leq e,
\]
where the inequality follows from $m/\delta \geq \ln \left( e/\varepsilon \right)$. Here we can also observe that $t^\star \leq \delta$. We now have
\[
h(t^\star) = \frac{e - \varepsilon}{\delta} (m - c - t^\star) + (e - \varepsilon) \geq \frac{e - \varepsilon}{\delta} (m - c)
,\]
where the inequality can be derived from  $t^\star \leq \delta$.
\end{proof}

By these lemmas, we can conclude that $(w, u, v)$ is a feasible solution to \eqref{prog:lbound-rand-dual2} whose objective value is at least $\frac{e - \varepsilon}{\delta} (m - c)$. We have argued that, by normalizing every variable by $\sum_{j = 1}^m b_j v_j$, we can derive a feasible solution to \eqref{prog:lbound-rand-dual1}. The next lemma indicates that $\sum_{j = 1}^m b_j v_j$ is sufficiently small.
\begin{lem}
We have $\sum_{j = 1}^m b_j v_j \leq (m/\delta) \cdot e^{1/\delta}$.
\end{lem}
\begin{proof}
Observe that
\begin{align*}
\sum_{j = 1}^m b_j v_j & = \sum_{j = 1}^m e^{j/\delta} \int_{(j-1)/\delta}^{j/\delta}e^{-z} dz \\
& = e^{1/\delta} \sum_{j = 1}^m e^{(j-1)/\delta} \int_{(j-1)/\delta}^{j/\delta}e^{-z} dz \\
& \leq e^{1/\delta} \sum_{j = 1}^m \int_{(j-1)/\delta}^{j/\delta} e^z \cdot e^{-z} dz \\
& = \frac{m}{\delta} \cdot e^{1/\delta}.
\end{align*}
\end{proof}

Therefore, we can obtain a solution feasible to \eqref{prog:lbound-rand-dual1} whose objective value is at least
\begin{equation}\label{e:finalvv}
\frac{(e - \varepsilon) \cdot (m/\delta - c/\delta)}{(m/\delta) \cdot e^{1/\delta}}.
\end{equation}
Given $\varepsilon$, we first fix $c/\delta$ as some constant satisfying $c/\delta\geq- \ln(e - \varepsilon) - \ln \ln \left(e/(e - \varepsilon)\right)$. Then we choose $\delta$ to be sufficiently large, which in turn determines $c$. After this, we choose $m$ to be sufficiently large. This shows that Equation~\eqref{e:finalvv} can be made arbitrarily close to $(e-\varepsilon)$. Now the desired conclusion follows from the weak duality of LP, completing the proof of Theorem~\ref{thm:lbound-rand}.
\end{proof}

\subsection{Deferred Proofs in Section~\ref{sec:lbound-det}}
\begin{lem}
Let $f:(0,1)\to\mathbb{R}_+$ be a continuous function. Suppose that, for all $\varepsilon>0$ and $\lambda \in (0,1)$, the robustness of any deterministic $(1+\lambda)$-consistent algorithm for the $(K,\beta)$-continuum button problem must be strictly greater than $f(\lambda)-\varepsilon$. Then, for all $\varepsilon>0$ and $\lambda \in (0,1)$, the robustness of any deterministic $(1+\lambda)$-consistent algorithm for the regular button problem must also be strictly greater than $f(\lambda)-\varepsilon$.
\end{lem}
\begin{proof}
Suppose towards contradiction that, for some $\lambda \in (0,1)$ and $\varepsilon > 0$, there exists a deterministic $(1+\lambda)$-consistent $(f(\lambda)-\varepsilon)$-robust algorithm $\A$ for the regular button problem. Without loss of generality, assume that $\varepsilon < f(\lambda)$. By choosing a sufficiently small $\varepsilon'>0$, we can satisfy $\varepsilon' < \varepsilon / 3$, $\lambda+\varepsilon' \in (0,1)$, and $f(\lambda + \varepsilon') > f(\lambda) - \varepsilon / 3$. Let $F:=\max\{2,\sup_{\lambda \in (0,1)} f(\lambda) \}$. We choose $\delta\in (0, (\varepsilon' \beta)/(FK)]$ so that $(m-1)/\delta$ becomes an integer.

Now consider the following algorithm for the $(K,\beta)$-continuum button problem. Let $b$ be the given price function and $\Jpred \in [1,m]$ be the prediction. The algorithm constructs an instance of the regular button problem by creating $((m-1)/\delta + 1)$ buttons whose values are $b((i-1) \delta +1)$ for $i=1,\cdots,(m-1)/\delta + 1$. The algorithm then internally executes $\A$ on this constructed instance, to which a prediction of $\tilde{J} := \ceil{(\Jpred -1)/\delta}+1$ is given. Each time $\A$ clicks a button, say $b_k$, the algorithm clicks the corresponding button $(k-1)\delta+1$ in the continuum problem. When the algorithm clicks a target, it reports to $\A$ that the last button was a target and terminates. Note that the total cost incurred by the algorithm is exactly equal to that by $\A$.

We claim that this algorithm is $(1+\lambda+\varepsilon')$-consistent and $(f(\lambda) - (2\varepsilon)/3)$-robust for the $(K,\beta)$-continuum button problem. Since $f(\lambda)-(2\varepsilon)/3 < f(\lambda+\varepsilon')-\varepsilon/3$, this leads to contradiction. It remains to verify the consistency and robustness.

Suppose we run the above algorithm when the first target $J$ of the continuum problem is equal to $\Jpred$. Note that the algorithm terminates when $\A$ clicks a button with index $\tilde{J}$ or higher. This means that the prediction $\tilde{J}$ given to $\A$ is also accurate (for the regular button problem). From the $(1+\lambda)$-consistency of $\A$, the total cost incurred by the algorithm is at most
\begin{align*}
 (1+\lambda) b_{\tilde J} &= (1 + \lambda) b \left(\ceil{(\Jpred-1)/\delta} \cdot \delta + 1 \right)\\
&\leq (1 + \lambda) (b(\Jpred) + K\delta)\\
&\leq(1+\lambda+\varepsilon') b(\Jpred),
\end{align*}
where the first inequality follows from the $K$-Lipschitz continuity of $b$ and the second from the monotonicity of $b$ and the choice of $\varepsilon'$. This shows the $(1+\lambda+\varepsilon')$-consistency of the algorithm.

Now we verify the robustness. Let $J$ be the first target button of the continuum problem. The algorithm terminates when $\A$ clicks a button with index $\ceil{(J -1)/\delta} +1$ or higher. Hence, from the $(f(\lambda)-\varepsilon)$-robustness of $\A$, the total cost incurred by the algorithm is at most
\begin{align*}
 (f(\lambda)-\varepsilon) b \left(\ceil{(J-1)/\delta} \cdot \delta + 1 \right) &\leq (f(\lambda)-\varepsilon) (b(J) + K\delta)\\
&\leq (f(\lambda)-\varepsilon+\varepsilon') b(J),
\end{align*}
where the first inequality again follows from the $K$-Lipschitz continuity of $b$ and the second from the monotonicity of $b$ and the choice of $\varepsilon'$. Note that $f(\lambda)-\varepsilon+\varepsilon' <f(\lambda)- (2\varepsilon)/3$, showing the desired robustness of the algorithm.
\end{proof}

\subsubsection{Lower Bound on Competitivenss} \label{app:lbound-det-comp}
Before presenting our trade-off between consistency and robustness for deterministic learning-augmented algorithms, let us first show that no deterministic algorithm for the $(K, \beta)$-continuum button problem can have a competitive ratio strictly better than $4$. Let $m$ be a sufficiently large number and $b(j) = j$ for $j \in [1, m]$. Note that $b$ is $1$-Lipschitz continuous with $b(1) = 1 > 0$.

Fix any deterministic algorithm for the continuum button problem. Let $n$ be the number of buttons the algorithm clicks; for any $i = 1, \cdots, n$, let $x_i$ be the button that the algorithm clicks in the $i$-th round. Note that, by the choice of $b$, the price of this button is also $x_i$. Moreover, we can without loss of generality assume that $x_1 < x_2 < \cdots < x_n$.

In order for the algorithm to be competitive, it should be competitive for any $J \in \{ 1, x_1 + \varepsilon, x_2 + \varepsilon, \cdots, x_{n - 1} + \varepsilon \}$ where $\varepsilon$ is an infinitesimal.
We therefore have the following infinite sequence of constraints as follows: for any integer $i = 0, 1, 2, \cdots, n - 1$,
\begin{equation} \label{eq:lbound-det-cons}
\sum_{j = 1}^{i+1} x_j \leq \gamma x_i,
\end{equation}
where $\gamma$ denotes the competitive ratio of the algorithm and $x_0 := 1$.

Let $\{a_i\}$ be an infinite sequence defined recursively as follows:
\[
a_1 = \gamma - 1 \text{ and } a_i = \gamma - \frac{\gamma}{a_{i - 1}} \text{ for } i \geq 2.
\]
Note that this infinite sequence is depends only on $\gamma$.

We show by induction that, for any $i = 1, \cdots, n-1$, we have
\begin{equation} \label{eq:lbound-det-lb}
x_{i + 1} \leq a_i x_i \text{ and } \sum_{j = 1}^{i + 1} x_j \geq \frac{\gamma}{a_i} \cdot x_{i + 1}.
\end{equation}
For $i = 1$, it is easy to see that $x_2 \leq (\gamma - 1) x_1$ due to Equation~\eqref{eq:lbound-det-cons} with $i = 1$. Moreover,
\[
x_ 1 + x_2 \geq \frac{1}{\gamma - 1} \cdot x_2 + x_2 = \frac{\gamma}{a_1} \cdot x_2,
\]
where the first inequality comes from $x_2 \leq a_1 x_1$, showing the second half of the statement. For $i \geq 2$, we have
\[
x_{i + 1} \leq \gamma x_i - \sum_{j = 1}^i x_j \leq \left( \gamma - \frac{\gamma}{a_{i - 1}} \right) x_i = a_i x_i
\]
where the first inequality comes from Equation~\eqref{eq:lbound-det-cons} and the second is derived from the induction hypothesis. We also have
\begin{align*}
\sum_{j = 1}^{i+1} x_j & \geq \frac{\gamma}{a_{i - 1}} \cdot x_i + x_{i + 1} \\
& \geq \frac{\gamma}{a_{i - 1}} \cdot \frac{1}{a_i} \cdot x_{i + 1}  + x_{i + 1} \\
& = \left( \frac{\gamma}{a_{i - 1}} \cdot \frac{1}{\gamma - \frac{\gamma}{a_{i - 1}}} + 1 \right) x_{i + 1}  \\
& = \frac{\gamma}{a_i} \cdot x_{i + 1},
\end{align*}
from the recurrence relation of $\{a_i\}_i$. This completes the proof of the claim.
 
The next two lemmas exhibit useful properties of $\{a_i\}_i$. These are crucial in proving the lower bound on competitive ratio for deterministic algorithms.
\begin{lem}\label{lem:lbound-conv}
If the sequence $\{a_i\}_i$ is convergent, we have $\gamma \geq 4$.
\end{lem}
\begin{proof}
Let $\alpha := \lim_{i \to \infty} a_i$. From the recurrence relation, we can obtain
\[
\alpha = \gamma - \frac{\gamma}{\alpha} \iff \frac{\alpha^2 - \gamma \alpha + \gamma}{\alpha} = 0.
\]
In order to have a real solution, we should have $\gamma^2 - 4 \gamma = \gamma (\gamma - 4) \geq 0$.
\end{proof}

\begin{lem} \label{lem:lbound-diver}
If $1 \leq \gamma < 4$, there exists a nonpositive element in the sequence $\{a_i\}_i$.
\end{lem}
\begin{proof}
Suppose towards contradiction that $\{a_i\}_i$ is all positive. Observe that, for any $i \geq 1$,
\[
a_i - a_{i + 1} = a_i - \left( \gamma - \frac{\gamma}{a_i} \right) = \frac{a_i^2 - \gamma a_i + \gamma}{a_i} > 0,
\]
where the inequality holds since $1 \leq \gamma < 4$, implying that the sequence is strictly decreasing. This implies that $\{a_i\}_i$, leading to contradiction from Lemma~\ref{lem:lbound-conv}.
\end{proof}

By choosing a sufficiently large $m$, we can make $n$ be arbitrarily large. By Lemma~\ref{lem:lbound-diver}, if $\gamma < 4$, there exists a nonpositive element in $\{a_i\}$. However, in that case, the algorithm cannot satisfy Equation~\eqref{eq:lbound-det-lb}. Hence, it is required to have $\gamma \geq 4$, completing the proof.

\subsubsection{Trade-off between Consistency and Robustness} \label{app:lbound-det-pred}
We are now ready to prove Theorem~\ref{thm:lbound-detpred}. We are given a constant $\lambda \in (0, 1)$. Let us bring the same instance where $m$ is a sufficiently large number and $b(j) = j$ for $j \in [1, m]$. Let the prediction $\Jpred \leq m$ be also sufficiently large. Fix any deterministic algorithm that utilizes the prediction $\Jpred$. As we want to guarantee the algorithm to be $(1 + \lambda)$-consistent, the algorithm must have incurred at most $\lambda \cdot \Jpred$ before it clicks button with index $\Jpred$ or higher. Let $n$ be the number of buttons the algorithm has clicked before it clicks button with index $\Jpred$ or higher. We thus obtain the following constraint on the algorithm:
\begin{equation} \label{eq:lbound-detpred-cons}
\sum_{j = 1}^n x_j \leq \lambda \cdot \Jpred,
\end{equation}
where $x_j$ again denotes the price of the button clicked at the $j$-th round as in the previous section.

Meanwhile, we also want the algorithm to be robust. Therefore, the algorithm should satisfy Equation~\eqref{eq:lbound-det-cons} for $i = 1, 2, \cdots, n - 1$ where $\gamma \geq 4$ now represents the robustness ratio, along with an additional constraint capturing the situation when $J = x_n + \varepsilon$ while the algorithm incurs at least $\Jpred$ after clicking button~$x_n$. In other words, we have
\begin{equation} \label{eq:lbound-detpred-rob}
\sum_{i = 1}^n x_i +\Jpred \leq \gamma x_n.
\end{equation}

Observe that, from Equations~\eqref{eq:lbound-detpred-cons} and~\eqref{eq:lbound-detpred-rob}, we have
\[
\left(1 + \frac{1}{\lambda} \right) \cdot \sum_{j = 1}^n x_j \leq \gamma x_n.
\]
Combining this inequality with Equation~\eqref{eq:lbound-det-lb}, we can obtain
\[
\left( 1 + \frac{1}{\lambda} \right) \cdot \frac{\gamma}{a_{n - 1}} x_n \leq \gamma x_n,
\]
resulting in that
\[
1 + \frac{1}{\lambda} \leq a_{n - 1}.
\]
When $m$ and $\Jpred$ is sufficiently large, we also have a sufficiently large $n$. We can thus replace $a_{n - 1}$ with $\alpha := \lim_{i \to \infty} a_i$ in the above inequality. From the proof of Lemma~\ref{lem:lbound-conv}, it is easy to see that $\alpha = \frac{\gamma + \sqrt{\gamma^2 - 4\gamma}}{2}$. We can thus obtain, for $\lambda \in (0, 1)$,
\begin{align*}
1 + \frac{1}{\lambda} \leq \frac{\gamma + \sqrt{\gamma^2 - 4\gamma}}{2} & \implies 2 + \frac{2}{\lambda} - \gamma \leq \sqrt{\gamma^2 - 4\gamma} \\
& \implies \left(2 + \frac{2}{\lambda} - \gamma \right)^2 \leq \gamma^2 - 4\gamma \\
& \implies \frac{4}{\lambda^2} + \frac{4}{\lambda}(2 - \gamma) + 4 \leq 0 \\
& \implies \gamma \geq 2 + \lambda + \frac{1}{\lambda}.
\end{align*}
Together with Lemma~\ref{lem:lbound-conti}, this completes the proof of Theorem~\ref{thm:lbound-detpred}.

\end{document}